          [2001/04/25 1.1 (PWD)]
\documentclass[a4paper,twocolumn]{esapub} 
\usepackage{psfig}

\newcommand\la{\raisebox{-0.5ex}{$\,\stackrel{<}{\scriptstyle\sim}\,$}}
\newcommand\ga{\raisebox{-0.5ex}{$\,\stackrel{>}{\scriptstyle\sim}\,$}}
\newcommand{\lya}{\ifmmode {\rm Ly}\alpha \else Ly$\alpha$\fi}
\newcommand{\hii}{H~{\sc ii}}
\def\msun{\ifmmode M_{\odot} \else M$_{\odot}$\fi}
\def\msunyr{\ifmmode M_{\odot} {\rm yr}^{-1} \else M$_{\odot}$ yr$^{-1}$\fi}
\def\zsun{\ifmmode Z_{\odot} \else Z$_{\odot}$\fi}
\def\lsun{\ifmmode L_{\odot} \else L$_{\odot}$\fi}
\def\micron{$\mu$m}
\def\aap{A\&A}

\def\aj{AJ}
\def\apj{ApJ}

\def\mnras{MNRAS}

\title{Starbursts at $z \protect\ga 6$: from the present to Herschel and ALMA}
\author{Daniel Schaerer}
\affil{Observatoire de Gen\`eve,
51, Ch. des Maillettes, CH-1290 Sauverny, Switzerland} 
\author{Roser Pell\'o}
\affil{Laboratoire d'Astrophysique (UMR 5572),
Observatoire Midi-Pyr\'en\'ees,
14 Avenue E. Belin, F-31400 Toulouse, France}

\begin{document}


\maketitle

\begin{abstract}
The SED of the lensed $z=6.56$ galaxy HCM6A behind the cluster Abell 370 has
been analysed. We find clear indications for the presence of dust in
this galaxy, and we estimate the properties of its stellar populations
(SFR, age, etc.). From its estimated luminosity, $L \sim (1-4) \times 10^{11}
\lsun$, this galaxy ranks as a luminous infrared galaxy . 
This case is then used to examine the detectability of high-z galaxies with
Herschel and ALMA. 
It is evident that with the use of strong gravitational lensing
SPIRE/Herschel observations could provide very interesting information
on $z>5$ galaxies. 
Strong synergies between ground-based near-IR instruments on 8-10m class
telescopes and ELTs, Herschel, the JWST, and ALMA can be expected 
for the exploration of the first galaxies in the Universe.
\end{abstract}

\section{Introduction}
Considerable advances have been made in the exploration of the high-z
Universe during the last decade, starting with the discovery and
detailed studies of redshift $z\sim 3$ galaxies (Lyman break
galaxies, LBGs), mostly from the pioneering work of Steidel
and collaborators (cf.\ Steidel et al.\ 2003),
reaching over the $z \sim$ 4--5 galaxies from different deep
multi-wavelength surveys, up to galaxies at $z \sim$ 6--7 close to the
end of the reionisation epoch of the Universe (e.g. Kodaira et al.\ 2002,
Hu et al.\ 2002, Cuby et al.\ 2003, Kneib et al.\ 2004, Stanway et
al.\ 2004, Bouwens et al.\ 2004).
To extend the present
searches beyond $z\ge$ 6.5 and back to ages where the Universe was
being re-ionized (cf.\ Fan et al.\ 2002), it is mandatory to move into
the near-IR bands.
This logical extension should allow to find and study the galaxies up to
$z \sim 10$ with ground based instruments and telescopes.

Other approaches could also be used to search for galaxies
during the re-ionisation epoch. Indeed, the recognition of 
a large negative k-correction of dust emission in the sub-mm and mm regime
which largely overcomes the effect of the inverse square law and cosmological
surface brightness dimming, and the existence of numerous dust rich
galaxies up to fairly high redshift, indicate that searches with high resolution
instruments as ALMA could locate high-z galaxies in an independent way
and probably even measure their redshift from fine structure lines
(see e.g.\ Blain et al.\ 2000, Guiderdoni et al.\ 1999).
However, to be feasible such searches obviously require that the target 
galaxies are chemically evolved and contain sufficient dust.
Although quasars up to the highest redshift currently known 
have revealed the presence of important quantities of dust (e.g. at $z=6.42$, 
Walter et al.\ 2004), little is known about dust in galaxies above 
$z \ga$ 5.
Actually most observations of Lyman break galaxies (LBG) at $z>4$ seem
to show relatively blue colors (and bluer ones than LBGs at lower $z$),
which possibly indicates less/little reddening in these objects
(see e.g.\ Stanway et al.\ 2004, Bouwens et al.\ 2004, review of Schaerer
2004).
Finding $z \gg 4$ galaxies with significant amounts of dust, and more 
generally studying the dust properties and content in high-z objects is 
therefore of great interest.

Here we first report on a recent analysis of a lensed $z=6.56$  starburst galaxy
indicating the presence of significant dust extinction (Schaerer \& Pell\'o 2004).
We then use this case to illustrate the detectability of such objects
with Herschel and ALMA, in particular in conjunction with 
strong gravitational lensing.

\begin{figure*}[htb]
\centerline{\psfig{figure=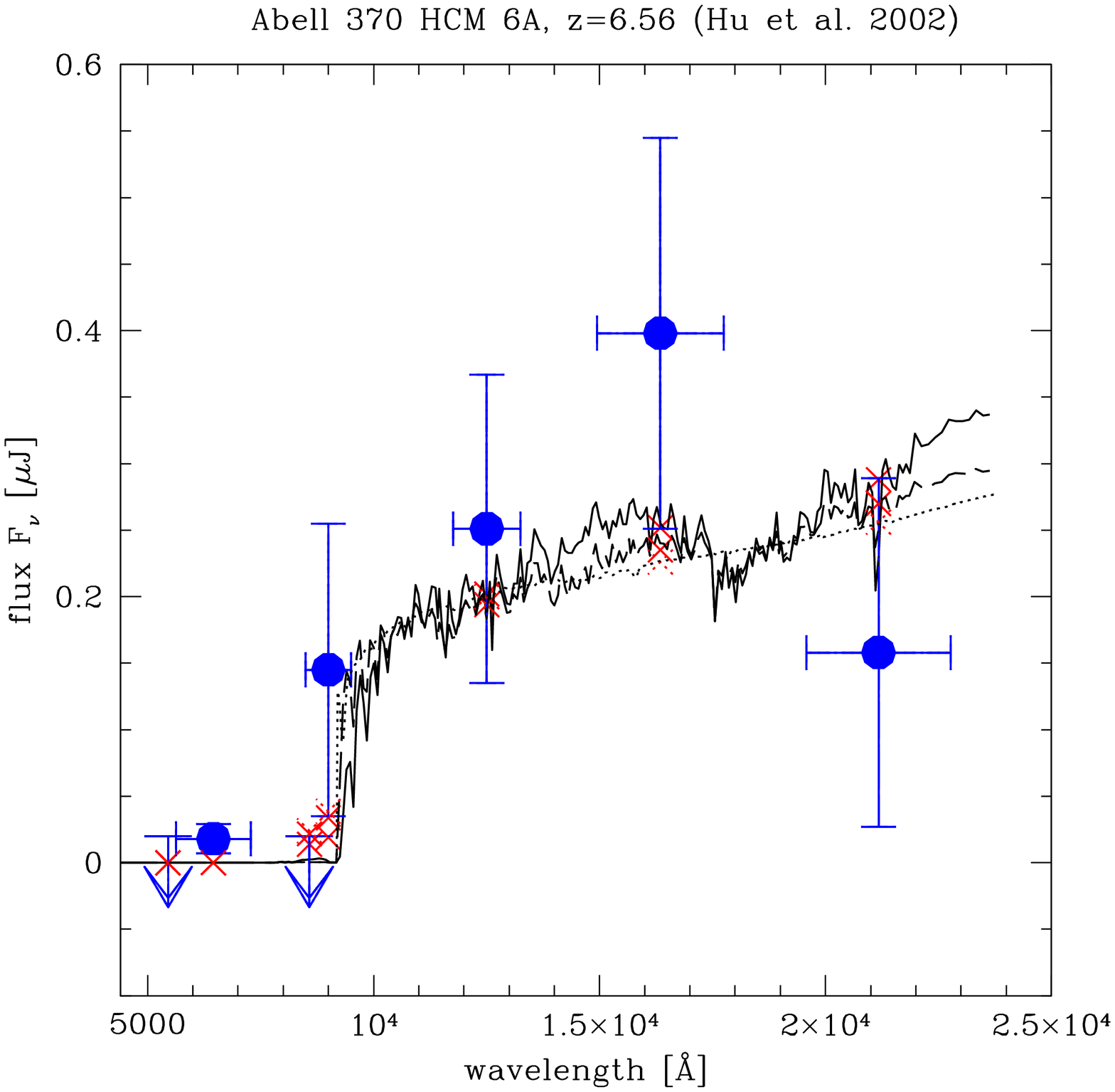,width=8.8cm}
	    \psfig{figure=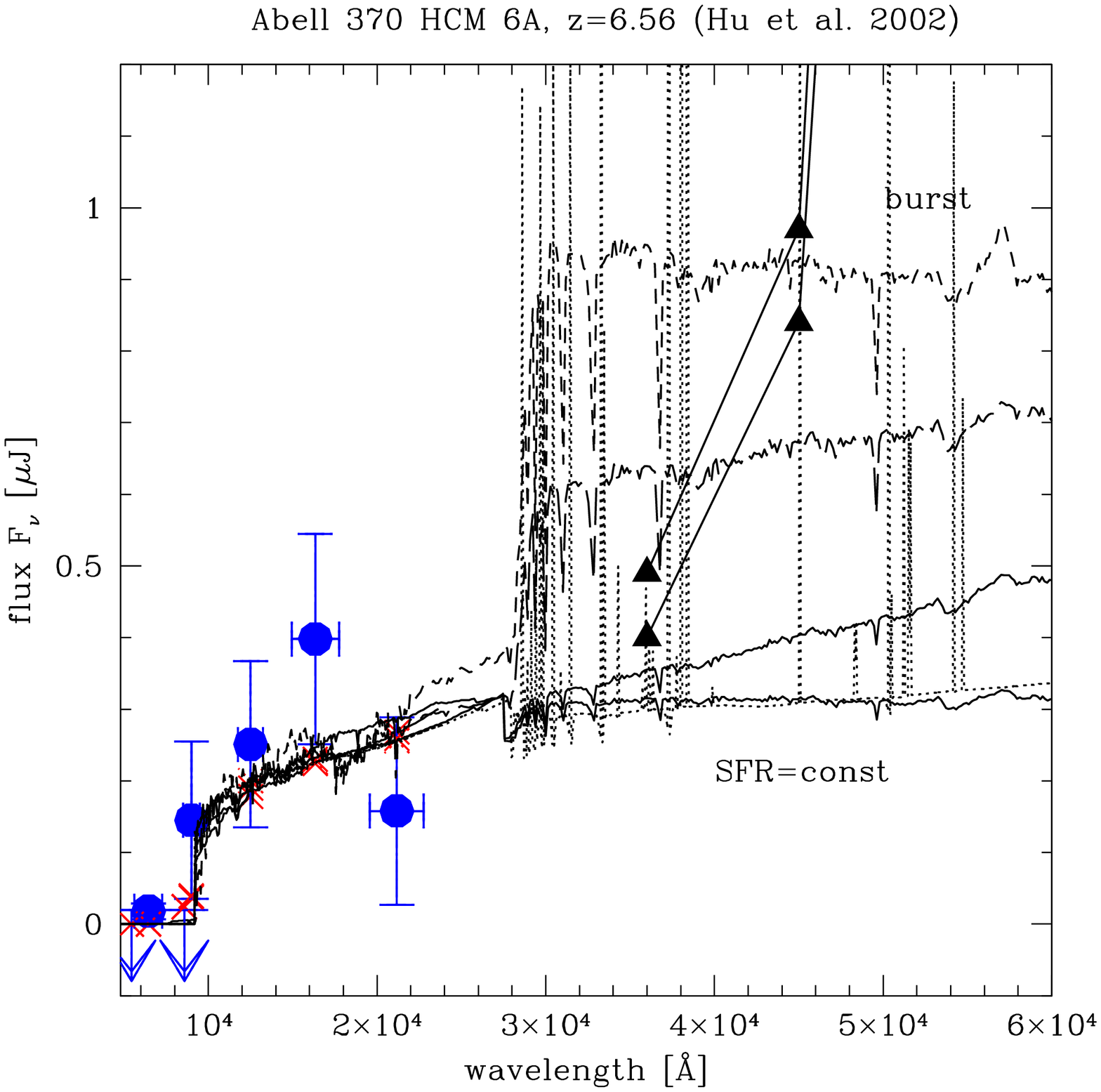,width=8.8cm}}
\caption{Best fits SEDs to the observations of Abell 370 HCM 6A.
The red crosses indicate the corresponding model broad band fluxes.
Solid lines show the best fit for a template from the BC+CWW group,
dotted from SB+QSO group, and dashed from the S03+ group 
{\bf Left:} Observed spectral range.
{\bf Right:} Predicted SED in Spitzer/IRAC domain for best fit models.
Dashed lines show the bursts from the BCCWW and S03+ template groups.
The dotted line is the spectrum of SBS 0335-052 from the SB+QSO group
with additional $A_V=1.$ The solid lines show best fits for constant
star formation using different extinction/attenuation laws (Calzetti 
starburst law versus SMC law). The solid triangles illustrate
the IRAC point-source sensitivity (1 $\sigma$) for low and medium
backgrounds excluding ``confusion noise''.}
\label{fig_sed_6a}
\end{figure*}

\begin{figure}[htb]
\centerline{
	    \psfig{figure=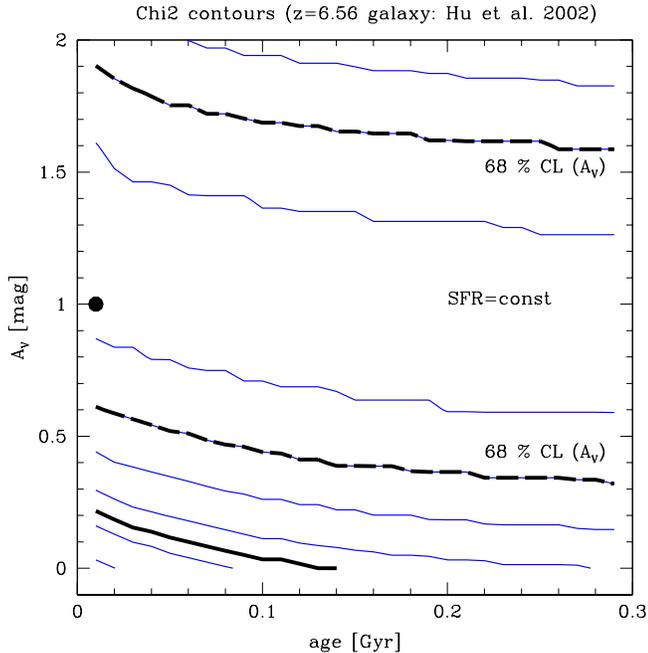,width=8.8cm}}
\caption{$\chi^2$ contour plots showing solutions in extinction -- age 
diagrams. The best solutions are indicate by the black dot.
Equidistant $\chi^2$ levels with a spacing of 0.5 are shown. The 2D 68\% 
confidence region (corresponding to $\Delta \chi^2 = 2.3$)
is delimited by the solid thick black line. The (1D) 68	\% confidence region
for $A_V$ ($\Delta \chi^2 = 1.$) at each given age is delimited by the  dashed thick black line.
Plot for solutions using a solar metallicity
burst template from the S03+ template group and the Calzetti attenuation law.
The solutions indicate an non-negligible extinction, but no constraint on age.
Discussion in text.
}
\label{fig_contour_6a}
\end{figure}

\section{Stellar populations and dust in a lensed $z=6.56$ starburst galaxy}
The lensed $z=6.56$ galaxy HCM6A was found by Hu et al.\ (2002)
from a narrow-band survey in the field of the lensing cluster Abell 370.
Its redshift is established from the broad-band SED including a strong
spectral break, and from the observed asymmetry of the detected emission
line identified as \lya.

We have recently analysed the SED of this object by means of quantitative
SED fitting techniques using a modified version of the {\em Hyperz} code of 
Bolzonella et al.\ (2000). 
The observed $VRIZJHK^\prime$ data are taken from Hu et al.\ (2002).
The gravitational magnification of the source is $\mu=4.5$
according to Hu et al.
The main free parameters of the SED modeling are
the spectral template, extinction, and the reddening law.
Empirical and theoretical templates including in particular 
starbursts and QSOs (SB+QSO templates), and predictions from synthesis models of 
Bruzual \& Charlot (BC+CWW group) and from Schaerer (2003, hereafter S03) are used. 

Overall the SED of HCM 6A (see Fig.\ 1)
is ``reddish'', showing an increase of the flux from Z to H
and even to K\footnote{The significance of a change of the SED slope
between JH and HK seems weak, and difficult to understand.}.
From this simple fact it is already clear 
qualitatively that one is driven towards stellar populations
with a) ``advanced'' age and little extinction or b) constant or 
young star formation  plus extinction.
However, for HCM6A a) can be excluded as no \lya\ emission would be expected 
in this case.

Quantitatively, the best solutions obtained for three ``spectral template groups''
are shown in the left panel of Fig.\ 1.
The solutions shown correspond to bursts of ages $\sim$ 50--130 Myr
and little or no extinction.
However, as just mentioned, solutions lacking young ($\la$ 10 Myr) massive stars can
be excluded since \lya\ emission is observed.
The best fit empirical SB+QSO template shown corresponds to the spectrum of 
the \hii\ galaxy SBS 0335-052 with an additional extinction of $A_V=1.$
On the basis of the present observations a narrow line (type II) AGN cannot be ruled out.
To reconcile the observed SED with \lya, a young population
e.g.\ such as SBS 0335-052 or constant SF is required. 
In any of these cases fitting the ``reddish'' SED requires
a non negligible amount of reddening.

To illustrate the typical range of possible results we show in
Fig.\ 2 $\chi^2$ a contour map and the corresponding confidence 
intervals for constant star formation, solar metallicity models
(S03+ template group) and reddened with the Calzetti law.
We see that for a given
age $A_V$ is typically $\sim$ 0.5--1.8 mag at the 68 \% confidence
level. For obvious reasons, no constraint can be set on the age 
since the onset of (constant) SF.
Hence, from the photometry of HCM 6A and from the presence of \lya\ 
we are led to conclude that this object must suffer from 
reddening with typical values of $A_V \sim 1.$ for a Calzetti
attenuation law.
A somewhat smaller extinction ($A_V \sim 0.4$) can be obtained if 
the steeper SMC extinction law of Pr\'evot et al.\ (1984) 
is adopted. From the present data it is not possible to distinguish
the different extinction/attenuation laws.

From the best fit constant SF models
we deduce an extinction corrected star formation rate of
the order of SFR(UV) $\sim$ 11 -- 41 \msunyr\ for a Salpeter
IMF from 1 to 100 \msun\ or a factor 2.55 higher for the often adopted
lower mass cut-off of 0.1 \msun.
For continuous SF over timescales $t_{\rm SF}$ longer than $\sim$ 10 Myr, the total
(bolometric) luminosity output is typically $\sim 10^{10}$ \lsun\ per unit
SFR (in \msunyr) for a Salpeter IMF from 1-100 \msun, quite independently of metallicity. 
The total luminosity  associated with the observed SF is therefore 
$L \sim (1-4) \times 10^{11} \lsun$, in the range of luminous infrared galaxies
(LIRG).
For $t_{\rm SF} \sim$ 10 Myr the estimated stellar mass is 
$M_\star \approx t_{\rm SF} \times SFR \sim (1-4) \times 10^8$ \msun. 
Other properties such as the ``\lya\ transmission'' can also
be estimated from this approach (see Schaerer \& Pell\'o 2004).

It is interesting to examine the SEDs predicted by the various
models at longer wavelengths, including the rest-frame optical
domain, which is potentially observable with the sensitive IRAC camera
onboard the Spitzer Observatory and other future missions.
In the right panel of Fig.\ 1 we plot 
again the 3 best fits.
We see that these solutions have fluxes comparable to or above 
the detection limit of IRAC/Spitzer 
\footnote{See {\tt http://ssc.spitzer.caltech.edu/irac/sens.html}}.
%
On the other hand the strongly reddened constant SF or young burst solutions
do not exhibit a Balmer break and are hence expected to show fluxes
just below the IRAC sensitivity at 3.6 \micron\ and significantly
lower at longer wavelengths.
As \lya\ emission is expected only for the reddened SEDs the
latter solutions are predicted to apply to HCM 6A.
If possible despite the presence of other nearby sources,
IRAC/Spitzer observations of HCM 6A down to the 
detection limit or observations with other future satellites 
could allow to verify our prediction and therefore provide an independent
(though indirect) confirmation of the presence of dust in this high-z
galaxy.

\section{$z \protect\ga 6$ starbursts: with Herschel and ALMA,
and now $\ldots$}

Let us now assume that starburst galaxies with dust exist at $z \ga 6$
and briefly examine their observability with facilities such as
Herschel and ALMA.
To do so we must assume a typical galaxy spectrum including the dust emission. 
For simplicity we here adopt the SED model by Melchior et al.\ (2001)
based on PEGASE.2 stellar modeling, on the D\'esert et al.\ (1990) dust
model, and including also synchrotron emission.
Their predicted SED for a galaxy with an SFR and/or total luminosity
quite similar to that estimated above for HCM6A is shown in Fig.\ 3.

\begin{figure}[htb]
\centerline{\psfig{figure=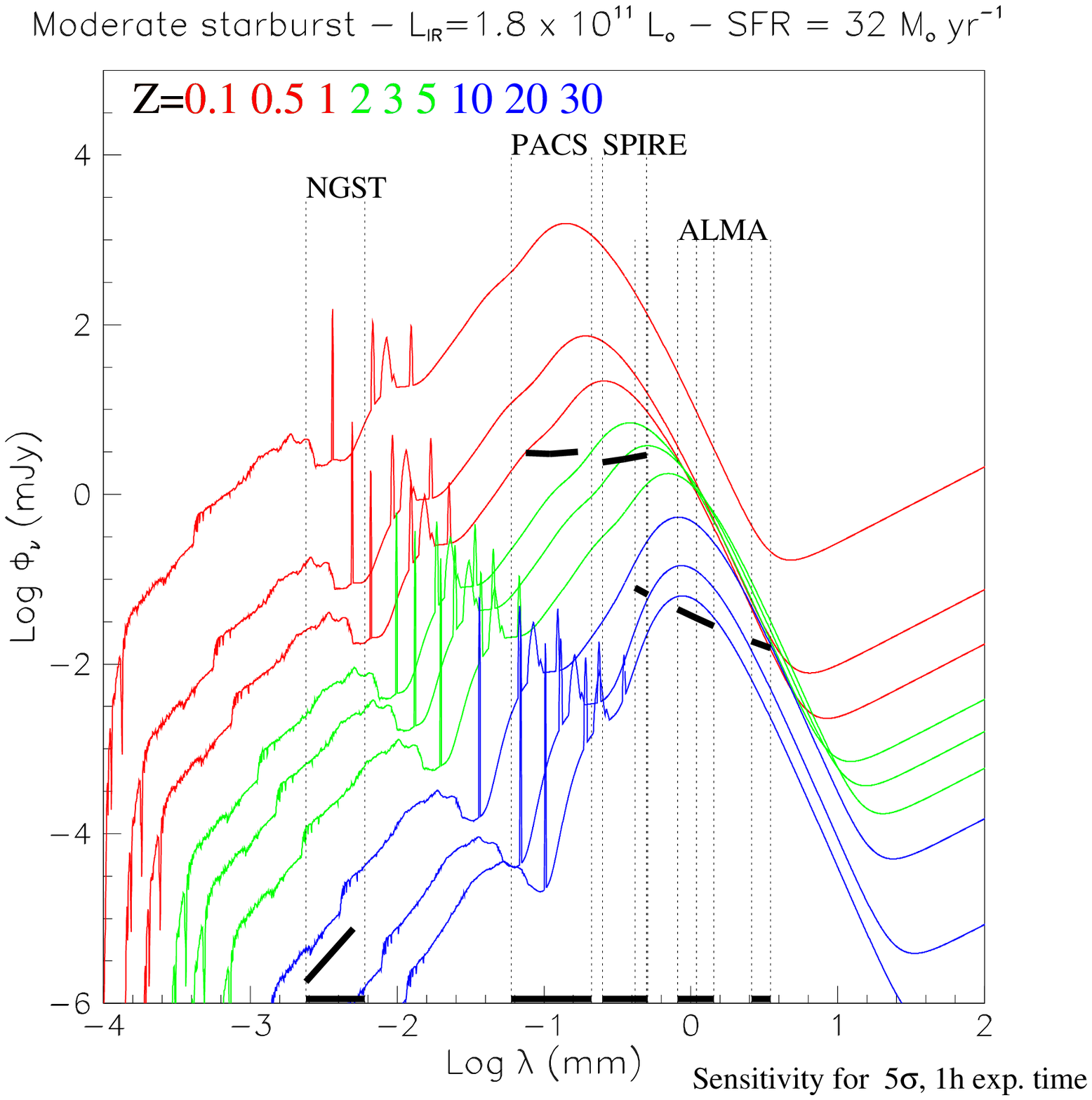,width=8.8cm}}
\caption{Predicted spectrum for a ``moderate'' starburst with SFR=32 \msunyr\
and $L \sim 1.8 \times 10^{11} \lsun$} placed at redshift $z=$0.1, 0.5, 1, 2, 3, 5, 10, 20,
and 30. The thresholds of the JWST (here NGST), PACS and SPIRE onboard Herschel,
and ALMA are also presented. Figure taken from Melchior et al.\ (2001) with kind permission
\label{}
\end{figure}

Figure 3 shows the exquisite sensitivity of ALMA in the various bands
allowing in principle an easy detection of such objects up to redshift $\sim$ 10 
or even higher!

On the other hand, with the sensitivity of PACS and SPIRE blank field 
observations of such an object are limited to smaller redshift ($z \la$ 1--4).
However, already with a source magnification of $\mu \sim$ 3--10 or more the
``template galaxy'' shown in Fig.\ 3 becomes observable with SPIRE
at $\sim$ 200--670 \micron.
In fact, such magnifications (and even higher ones) are not exceptional
in the central parts of massive lensing clusters. E.g.\ in our near-IR search 
conducted in two ISAAC fields ($\sim$ 2.5x2.5 arcmin$^2$) of two lensing clusters,
a fair number ($\sim$ 10--20) of $z \ga$ 6--7 galaxy candidates with $\mu \ga 5$
are found (Pell\'o et al.\ 2004; Richard et al.\ 2004, in preparation). 
More than half of them have actually magnifications $\mu \ga 10$!
Such simple estimates show already quite clearly the potential of
strong gravitational lensing to extend the horizon of SPIRE/Herschel observations
beyond redshift $z \ga 5$!

Obviously a more rigorous feasibility study must also address the following
issues: 
How frequent is dust present in high-z galaxies? and up to what redshift?
We now have some indications for dust in one lensed $z=6.56$ galaxy (see Section 2)
and of course in high-z quasars. But how general/frequent is this?
How typical is the SED adopted above? The long wavelength emission due to dust
depends on various parameters such as metallicity, the dust/gas ratio, geometry,
the ISM pressure etc. 
Furthermore spatial resolution and source confusion are key issues which
must be addressed and which should vary quite strongly between blank fields
and cluster environments.
Last, but not least, the field of view of the various instruments is determinant
for the efficiency with which high-z candidates can be found and studied.
Several of these isssues have already been partly addressed earlier
(cf.\ the 2000 Herschel conference proceedings of Pilbratt et al.\ 2001,
also Blain et al.\ 2002).

It is evident that various ground-based and space bourne facilities and
instruments will be used together to provide an optimal coverage
in wavelength, spatial resolution and field size, and to obtain imaging
as well as spectroscopy. 
Near-IR wide field imagers and near-IR multi-object spectrographs
on 8-10m class telescopes and later with ELTs will undoubtably ``team up''
with the JWST, Herschel and ALMA to explore the first galaxies
in the Universe and their evolution from the Dark Ages to Cosmic Reionisation.



{\small 
\begin{thebibliography}{}
\bibitem{} Blain, A.W., et al., 2000, \mnras, 313, 559
\bibitem{} Blain, A.W., et al., 2002, Physics Reports, 369, 111 
\bibitem{} Bolzonella, M., Miralles, J.-M., Pell\'o, R., 2000, \aap, 363, 476
\bibitem{} Bouwens, R.J., et al, 2004, \apj, in press [astro-ph/0409488]
\bibitem{} Cuby, J.-G., et al., 2003, \aap, 405, L19
\bibitem{} D\'esert et al., 1999, \aap, 237, 215
\bibitem{} Fan, X., et al., 2002, \aj, 123, 1247
\bibitem{} Guiderdoni, B., et al., 1999, in ``The Birth of Galaxies'', B. Guiderdoni, et al. (eds), 
	Editions Frontieres, [astro-ph/9902141]
\bibitem{} Hu, E.M., et al., 2002, \apj, 568, L75; Erratum: \apj, 576, L99
\bibitem{} Kneib, J.P., et al., 2004, \apj, 607, 697
\bibitem{} Kodaira, K., et al., 2003, PASJ, 55, L17
\bibitem{} Melchior, A.-L., et al., 2001, in ``The Promise of the Herschel Space Observatory'',
ESA SP-460, p.\ 467 [astro-ph/0102086]
\bibitem{} Pilbratt, G.L., et al., Eds., 2001, ``The promise of the Herschel Space Observatory'', ESA-SP 460
\bibitem{} Pell\'o, R., et al., 2004,  IAU Symposium No. 225, ``The Impact of Gravitational Lensing on Cosmology'', Y. Mellier and G. Meylan, Eds., [astro-ph/0410132]
\bibitem{} Schaerer, D., 2003, \aap, 397, 527
\bibitem{} Schaerer, D., 2004, in ``Starbursts: from 30 Doradus to Lyman break galaxies'', Eds. de Grijs, 
González Delgado,. ApSS, in press
\bibitem{} Schaerer, D., Pell\'o, R., 2004, A\&A, submitted
\bibitem{} Stanway, E., et al., 2004, MNRAS, submitted [astro-ph/0403585]
\bibitem{} 
     Steidel, C.~C., et al., 2003, \apj 592, 728
\bibitem{} Walter, F., et al., 2004, ApJ, 615, L17
\end{thebibliography}
}

\end{document}